\documentclass[final]{elsart}
\usepackage{ifpdf}
\journal{Optics Communications}
\usepackage{graphicx,amssymb,lineno}
\usepackage{amsmath}
\ifpdf
\usepackage[%
  pdftitle={Instructions for use of the document class
    elsart},%
  pdfauthor={Simon Pepping},%
  pdfsubject={The preprint document class elsart},%
  pdfkeywords={instructions for use, elsart, document class},%
  pdfstartview=FitH,%
  bookmarks=true,%
  bookmarksopen=true,%
  breaklinks=true,%
  colorlinks=true,%
  linkcolor=blue,anchorcolor=blue,%
  citecolor=blue,filecolor=blue,%
  menucolor=blue,pagecolor=blue,%
  urlcolor=blue]{hyperref}
\else
\usepackage[%
  breaklinks=true,%
  colorlinks=true,%
  linkcolor=blue,anchorcolor=blue,%
  citecolor=blue,filecolor=blue,%
  menucolor=blue,pagecolor=blue,%
  urlcolor=blue]{hyperref}
\fi

\makeatletter
\def\elsartstyle{%
    \def\normalsize{\@setfontsize\normalsize\@xiipt{14.5}}
    \def\small{\@setfontsize\small\@xipt{13.6}}
    \let\footnotesize=\small
    \def\large{\@setfontsize\large\@xivpt{18}}
    \def\Large{\@setfontsize\Large\@xviipt{22}}
    \skip\@mpfootins = 18\p@ \@plus 2\p@
    \normalsize
}
\@ifundefined{square}{}{\let\Box\square}
\makeatother

\begin{document}

\begin{frontmatter}
\title{Engineering two-mode squeezed states of cold atomic clouds with a superconducting stripline resonator}
\author{\corauthref{Li}Peng-Bo Li},
\corauth[Li]{Corresponding author.} \ead{lipengbo@mail.xjtu.edu.cn, Tel:
+86-29-82663714;Fax:+86-29-82667872}
\author{Fu-Li Li}
\address{MOE Key Laboratory for Nonequilibrium Synthesis and Modulation of Condensed Matter,\\
Department of Applied Physics, Xi'an Jiaotong University, Xi'an
710049, China}

\begin{abstract}
A scheme is proposed for engineering two-mode squeezed states of two separated cold atomic clouds positioned near the surface of a superconducting stripline resonator. Based on the coherent magnetic
coupling between the atomic spins and a stripline resonator mode,
the desired two-mode squeezed state can be produced via precisely control on the dynamics of the system. This scheme may be used to realize scalable on-chip quantum networks with cold atoms coupling to stripline resonators.
\end{abstract}

\end{frontmatter}

Two-mode squeezed states, which exhibit Einstein-Podolsky-Rosen entanglement, play a significant role in realizing  quantum information protocols and investigating quantum nonlocality with continuous-variables [1]. Since entangled distant atomic ensembles are the building blocks for scalable quantum networks [2], several proposals have been given to produce two-mode squeezed states of distant
atomic ensembles with cavity QED system [3,4]. Recently, hybrid systems consisting of ensembles of atomic or molecular system and
superconducting transmission line resonators have been intensively investigated [5-21]. Solid-state superconducting quantum circuits are
believed to have the advantages of integration and scaling on a
chip [22]. Ultracold atoms are very attractive candidates for storing and processing quantum information, in view of their long coherence times and the well-developed techniques for detecting and manipulating the ground electronic
(hyperfine) states. To realize scalable on-chip quantum networks or other quantum information applications with hybrid systems of cold atoms and stripline resonators, it is desirable to design a scheme that can generate two-mode squeezed states of separated cold atomic clouds in a controllable way.

In this work, utilizing the strong magnetic coupling of cold atoms to a superconducting transmission
line resonator [8], we propose a hybrid scheme to engineer two-mode squeezed states of two separated cold atomic clouds. We show that, under certain conditions the coupled system of two atomic
spin modes and the cavity mode can behave as three coupled harmonic
oscillators with controllable coefficients. Through coherent
control on the dynamics of the system, at some instants the
cavity mode is decoupled from the atomic modes, leaving
the latter in a two-mode squeezed state. Compared to other proposals based on electric-dipole coupling between
atoms and fields in cavity QED [3,4], this scheme,
utilizing strong magnetic coupling of atomic spins to cavity
mode [8], has the advantage of being immune to charge
noise and could constitute qubits with much longer coherence
times. Moreover, this protocol, which is robust against the effect of thermal photons, is scalable and much easier to be integrated on a chip.
With the technology of atomchip and circuit
QED, our study may open promising perspectives for the implementation
of on-chip quantum networks.

The system we consider consists of two separated atomic clouds trapped above the surface of a superconducting transmission
line resonator, as sketched in Fig. 1.  
\begin{figure}[h]
\centering
\includegraphics[bb=110 504 489 660,totalheight=1.3in,clip]{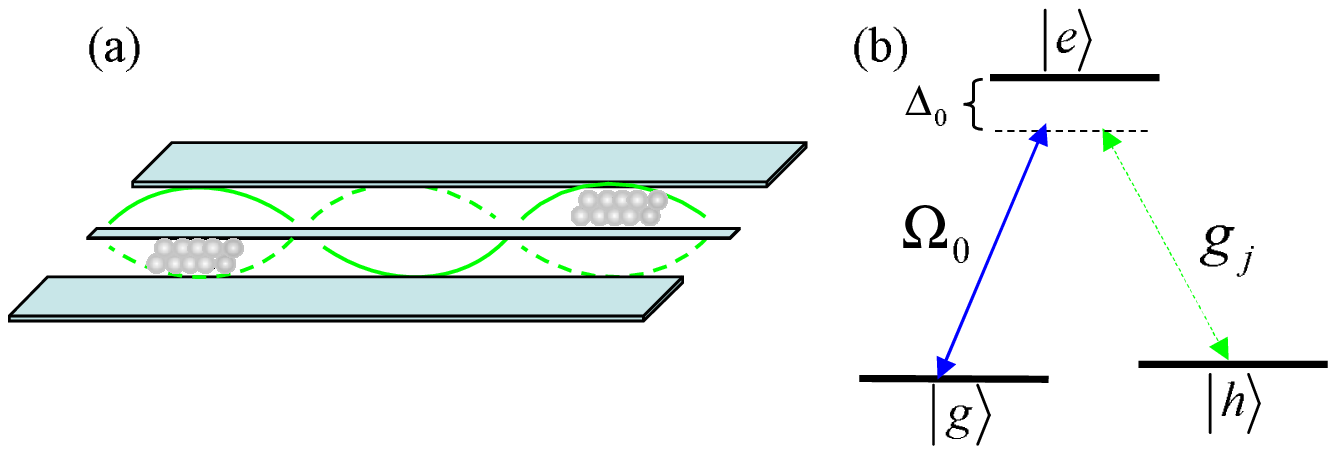}
\caption{(a) Two clouds of ultracold atoms positioned above the surface of
a superconducting transmission line resonator. (b) Atomic level structure with couplings to the cavity mode and driving field.
The atoms in ensemble 1 and 2 are initially cooled in the ground states $\vert \textbf{0}\rangle_1$ and $\vert \textbf{0}\rangle_2$, respectively.}
\end{figure}
Each atomic cloud contains $N_i(i=1,2)$ atoms. The transmission
line resonator consists of three conducting stripes: the central
conductor plus two ground planes. The electromagnetic field
of the resonator is confined near the gaps between the
conductor and the ground planes. Employing a variety of macroscopic
electrostatic traps or other atom chip technology, clouds of cold $^{87}$Rb atoms can
be trapped close to the resonator [23]. Each atom has the level structure of three-level $\Lambda$ configuration, with $\vert g\rangle$ and $\vert h\rangle$ the ground states, and $\vert e\rangle$ the excited state. The dominant interactions with a microwave field for Rb atoms cooled in the ground state are the magnetic
dipole transitions between the atomic hyperfine states of the $5S_{1/2}$ ground state. We choose  $\vert F=1,m_F=-1\rangle=\vert g\rangle$, $\vert F=1,m_F=1\rangle=\vert h\rangle$, and $\vert F=2,m_F=0\rangle=\vert e\rangle$ [23]. Assume that initially the atoms in the first ensemble are cooled in the ground state $\vert g\rangle$, and those in the second ensemble $\vert h\rangle$.
The
classical field (frequency $\omega_0$) drives the transition $\vert g\rangle\leftrightarrow \vert e\rangle$ with Rabi frequency $\Omega_0$, while the cavity mode (frequency $\nu_0$) couples the transition $\vert h\rangle\leftrightarrow \vert e\rangle$ with the coupling constant $g_j$ in the $j$th ensemble. To cancel the additional ac-Stark shifts
to the states $\vert g\rangle$ and $\vert h\rangle$, we need two virtually excited
states $\vert r\rangle$ and $\vert s\rangle$ \cite{24}. The state $\vert r\rangle$ is virtually excited from the ground state $\vert g\rangle$ by another
largely detuned classical field (frequency $\omega_1$) with Rabi frequency $\Omega_1$, while the cavity mode virtually couples the states $\vert h\rangle $ and $\vert s\rangle$ with coupling constant $\lambda_j$. The corresponding detunings for the related transitions are $\Delta_0=\omega_{e g}-\omega_0=\omega_{eh}-\nu_0$, $\Delta_r=\omega_{rg}-\omega_1$, and $\Delta_s=\omega_{sh}-\nu_0$, where $\omega_{eg},\omega_{eh},\omega_{rg},\omega_{sh}$ are the transition frequencies for the atoms. In the interaction
picture, the Hamiltonian describing the atom-field interaction
is (let $\hbar=1$)
\begin{eqnarray}
\label{H1}
H_I
&=&\sum_{j=1,2}\{\sum_{n=1}^{N_j}[\Omega_0\vert e_n^j\rangle\langle g_n^j\vert e^{i\Delta_0t}+
g_j\hat{a}\vert e_n^j\rangle\langle h_n^j\vert e^{i\Delta_0t}\nonumber\\
&&+\Omega_1\vert r_n^j\rangle\langle g_n^j\vert e^{i\Delta_rt}+\lambda_j\hat{a}\vert s_n^j\rangle\langle h_n^j\vert e^{i\Delta_st}]\}+\mbox{H.c.},
\end{eqnarray}
where $\hat{a}$ is the annihilation operator for the cavity mode. We consider dispersive
detuning
$|\Delta_0|,|\Delta_r|,|\Delta_s|\gg|\Omega_0|,|\Omega_1|,|g_j|,|\lambda_j|$. Therefore,
the excited states can be adiabatically eliminated. Then the interaction Hamiltonian describing the coupling of the cold atoms to the cavity mode is
\begin{eqnarray}
H_{I}&=&\sum_{j=1,2}\left[\left(\frac{\Omega_0^2}{\Delta_0}+\frac{\Omega_1^2}{\Delta_r}\right)\sigma^j_{gg}+\left(\frac{g_j^2}{\Delta_0}+\frac{\lambda_j^2}{\Delta_s}\right)\sigma^j_{hh}\hat{a}^\dag\hat{a}\right]\nonumber\\
&&+(\beta_2\hat{c}_2+\beta_1\hat{c}_1^\dag)\hat{a}^\dag+\mbox{H.c}.,
\end{eqnarray}
where $\sigma^j_{gg}=\sum_{n=1}^{N_j}\vert g_n^j\rangle\langle g_n^j\vert$, $\sigma^j_{hh}=\sum_{n=1}^{N_j}\vert h_n^j\rangle\langle h_n^j\vert$, $\beta_i=\frac{\sqrt{N_i}\Omega_0g_i}{\Delta_0}$, $\hat{c}_1^\dag=1/\sqrt{N_1}\sum_{n=1}^{N_1}\vert h^1_n\rangle\langle g_n^1\vert$, and $\hat{c}_2^\dag=1/\sqrt{N_2}\sum_{n=1}^{N_2}\vert g^2_n\rangle\langle h^2_n\vert$.
To appropriately choose the parameters, we assume
that $\frac{\Omega_0^2}{\Delta_0}+\frac{\Omega_1^2}{\Delta_r}=0$ and $\frac{g_j^2}{\Delta_0}+\frac{\lambda_j^2}{\Delta_s}=0$.
For weak
excitations of the atoms, the operator $\hat{c}_i$ obeys
approximate harmonic oscillator commutation relations
$[\hat{c}_i,\hat{c}_i^\dag]\simeq1$. In such a case, we can describe the
atomic excitations in the $i$th ensemble as a set of harmonic oscillator states
$\vert\textbf{0}\rangle_{1}=\vert g_1g_2...g_N\rangle_1$, $\vert\textbf{0}\rangle_{2}=\vert h_1h_2...h_N\rangle_2$,
$\vert\textbf{1}\rangle_i=\hat{c}_i^\dag\vert\textbf{0}\rangle_i$,
etc. Assume $\beta_1=i\xi_1,\beta_2=i\xi_2,|\xi_2|>|\xi_1|$, and introduce $\Theta=\sqrt{|\xi_2|^2-|\xi_1|^2}$.
Then we obtain the effective Hamiltonian
\begin{eqnarray}
\label{H}
H_I&=&H_1+H_2\nonumber\\
&=&i\xi_1\hat{c}_1^\dag\hat{a}^\dag-i\xi_1^*\hat{c}_1\hat{a}+i\xi_2\hat{c}_2^\dag\hat{a}-i\xi_2^*\hat{c}_2\hat{a}^\dag.
\end{eqnarray}
The above Hamiltonian shows that the ensemble excitations and the cavity mode represent a system of three coupled
harmonic oscillators with controllable coefficients. Hamiltonian $H_1$ describes simultaneous creation or annihilation of
an atomic spin in ensemble 1 and a photon, while $H_2$ describes
the exchange of excitation quanta between the collective
atomic spin-waves and cavity mode.

The Hamiltonian $H_I$ commutates with the
constant of motion $\hat{N}=\hat{c}_2^\dag\hat{c}_2-\hat{c}_1^\dag\hat{c}_1+\hat{a}^\dag\hat{a}$. So if the system starts from the state
$\vert 0\rangle_c\vert\textbf{0}\rangle_1\vert\textbf{0}\rangle_2$, with $\vert n\rangle_c$ is the Fock state for the cavity mode, then we have
$\hat{N}=0$ at any time during the evolution. The time evolution of
the operators in the Heisenberg representation is given
by [25]
\begin{subequations}
\begin{eqnarray}
\hat{c}_1(t)&=&\frac{\xi_1}{\Theta}\hat{a}^\dag(0)\sin\Theta
t-\frac{\xi_1\xi_2}{\Theta^2}[1-\cos\Theta
t]\hat{c}_2^\dag(0)+\frac{1}{\Theta^2}[|\xi_2|^2\nonumber\\
&&-|\xi_1|^2\cos\Theta
t]\hat{c}_1(0),\\
\hat{c}_2(t)&=&\frac{\xi_2}{\Theta}\hat{a}(0)\sin\Theta
t+\frac{\xi_1\xi_2}{\Theta^2}[1-\cos\Theta
t]\hat{c}_1^\dag(0)-\frac{1}{\Theta^2}[|\xi_1|^2\nonumber\\
&&-|\xi_2|^2\cos\Theta
t]\hat{c}_2(0),\\
\hat{a}(t)&=&\hat{a}(0)\cos\Theta
t+\frac{1}{\Theta}[-\xi_2^*\hat{c}_2(0)+\xi_1\hat{c}_1^\dag(0)]\sin\Theta
t.
\end{eqnarray}
\end{subequations}
Generally these solutions describe tripartite entanglement between
cavity mode and collective spin excitations. One can find that
after half a period $T_\pi=\pi/\Theta$, the cavity mode is
decoupled from the collective atomic spin excitations. Moreover, the
two atomic spin modes exhibit quantum correlations, namely
\begin{subequations}
\label{E1}
\begin{eqnarray}
\hat{c}_1(T_\pi)=\frac{|\xi_1|^2+|\xi_2|^2}{\Theta^2}\hat{c}_1(0)-\frac{2\xi_1\xi_2}{\Theta^2}\hat{c}^\dag_2(0),\\
\hat{c}_2(T_\pi)=\frac{2\xi_1\xi_2}{\Theta^2}\hat{c}^\dag_1(0)-\frac{|\xi_1|^2+|\xi_2|^2}{\Theta^2}\hat{c}_2(0).
\end{eqnarray}
\end{subequations}
This result shows that if at $t=0$ the state of the system is
$\vert\Psi(0)\rangle=\vert \textbf{0}\rangle_1\vert \textbf{0}\rangle_2\vert 0\rangle_c$, then at $t=T_\pi$,
the two atomic modes are in the two-mode squeezed state
\begin{equation}
\label{EPR}
\vert\phi\rangle_a=\left(\frac{1-r^2}{1+r^2}\right)\sum_{n=0}^\infty[\frac{2r}{1+r^2}]^n\vert \textbf{n}\rangle_1\vert \textbf{n}\rangle_2,
\end{equation}
while the cavity mode has been decoupled from two atomic modes. The squeezing parameter is
$\epsilon=\tanh^{-1}(\frac{2r}{1+r^2})$, which is determined by
$r=|\xi_2/\xi_1|$, thus by the ratio $|\beta_2/\beta_1|$.
When the squeezed
state is generated at the time of $T_\pi$, we switch off the lasers and decouple the atomic clouds to the cavity. Then the squeezed state can be preserved.

It is known that the two-mode squeezed state exhibits EPR entanglement. To further quantify the squeezing property of the two atomic modes, we employ the two-mode relative number
squeezing parameter [26] $\zeta_{12}=\sigma^2(\hat{c}_1^\dag\hat{c}_1-\hat{c}_2^\dag\hat{c}_2)/(\langle\hat{c}_1^\dag\hat{c}_1\rangle+\langle\hat{c}_2^\dag\hat{c}_2\rangle)$, where $\sigma^2(X)=\langle X^2\rangle-\langle X\rangle^2$.
$\zeta_{12}$ taking the value of 0 signifies two-mode squeezing of the atomic spin modes, while $\zeta_{12}=1$ means the two atomic modes are in independent states. To calculate the parameter $\zeta_{12}$ under the effect of cavity losses, we employ the quantum master equation approach.  The evolution of
the system is governed by the following master equation
\begin{eqnarray}
\frac{d\rho}{dt}&=&-\frac{i}{\hbar}[H_I,\rho]+\kappa\hat{a}\rho\hat{a}^\dag-\kappa\{\hat{a}^\dag a,\rho\},
\end{eqnarray}
where $\kappa$ stands for the cavity photon decay rate.
\begin{figure}[h]
\centering
\includegraphics[bb=29 392 453 763,totalheight=2in,clip]{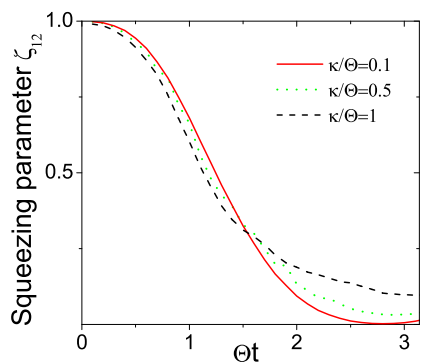}
\caption{Time evolution of the parameter $\zeta_{12}$ under $\kappa/\Theta=0.1,0.5,1$. Other parameters are chosen such that $|\beta_2/\beta_1|=1.5$.}
\end{figure}
In Fig. 2, we display the numerical results for the time evolution of the two-mode relative number
squeezing parameter for several values of the parameter $\kappa$ through the Monte Carlo wave function method. We see that nearly ideal squeezed states between the two atomic modes can occur at the instant $T_\pi=\pi/\Theta$ in the strong coupling regime $\Theta\gg\kappa$. In fact, even in the relatively weak coupling regime $\Theta\sim\kappa$, one can still obtain two-mode squeezing between the two atomic modes at the instant $T_\pi=\pi/\Theta$.

We now analyze the experimental feasibility of this proposal. As mentioned above, a promising
candidate for the atoms is $^{87}$Rb coupled to a stripline
resonator. With the help of atom-chip technology, ensembles of cold Rb atoms and a superconducting transmission line resonator could be integrated in a hybrid device on a single superconducting atomchip. The preparation of the initial atomic state can be accomplished using the well-developed optical pumping and adiabatic population transfer techniques. We assume modest atomic densities, so that atomic interactions can safely be neglected when they are in the ground state. In such a case, mechanical interaction between atoms can be avoided, and the internal degree of freedom of atoms can be decoupled from atomic motion. A typical ensemble of cold Rb atoms confined in an elongated trap on the atomchip possesses a transverse extension of about 1 $\mu$m and a length of up to several millimeters. In this situation, the variation of the microwave field of the resonator is neglectable on these length scales. If the atomic clouds are fixed parallel to the transmission line resonator, we can neglect the change in the microwave field over each atomic cloud and assume uniform coupling constants for all the atoms in each ensemble. For a transmission line cavity
with the strip-line length $L$ and effective dielectric constant
$\epsilon_r$, the frequency of the $m$th standing-wave mode of the
cavity is $\omega_c=\pi mc/L\sqrt{\epsilon_r}$. The coupling between
the atoms and the cavity mode can be estimated as
$g\sim D\sqrt{\frac{2\hbar\omega_c}{\pi^2h^2L}}$, where $D$ represents the dipole
matrix element and $h$ is the height for the atoms above the cavity
surface. For an atomic ensemble of $N\sim10^6-10^8$ $^{87}$Rb atoms trapped several $\mu$m above the surface of a stripline cavity, a collective coupling strength of $\sqrt{N}g/2\pi\sim400$
kHZ can be obtained, which dominates the cavity decay $\kappa/2\pi\sim7$ kHZ.
We choose the coupling strength for the photon-atom interaction in each ensemble as $g_1\sim g$, $g_2\sim1.5g$, and the detuning $\Delta_0\sim10g$.
For the hyperfine transition between $\vert F=1,m_F\rangle$ and  $\vert F=2,m_F\rangle$ states at a frequency of
$\nu_a=\omega_a/2\pi=6.83$ GHz, the frequencies of the classical Raman field and the cavity mode can be $\omega_0\sim\nu_a-\Delta_0$, $\nu_0\sim\nu_a-\Delta_0$.
With the given parameters, one can obtain the angular frequency
$\Theta/2\pi\sim90$ kHZ, and the time to prepare the squeezed state $T_{\pi}\sim5\mu$s (less than $1/\kappa\sim25\mu$s). The readout this squeezed state can be accomplished by matter-light state mapping techniques. Employing another electronic excited state to establish two distinct Raman transition channels via optical fields between the two ground states, the states of ensembles 1
and 2 can be transferred to propagating light outputs, which can be detected via the standard homodyne detection.

It is worth emphasizing that, though the cavity mode plays a fundamental role in establishing the
entanglement between the atomic modes, its initial state does not affect
the efficiency of the process. Therefore, if the cavity mode is in a thermal distribution initially, the above discussion still holds. So this scheme is robust against the effect of thermal photons, which is a distinct feature compared to the scheme proposed in Ref.[3]. Compared to other protocols for producing two-mode squeezed states of two distant atomic clouds based on electric-dipole coupling between atoms and fields in cavity QED, this
scheme, utilizing strong magnetic coupling of atomic spins to cavity mode, has the advantage of being immune
to charge noise and could constitute qubits with much
longer coherence times. Moreover, this proposal can be extended to implement scalable on-chip quantum networks, as entangled distant atomic ensembles are the building blocks for quantum
networks.

To conclude, we have proposed an efficient scheme to generate two-mode squeezed states of two separated cold atomic clouds coupled to a  superconducting transmission line resonator. It is based on the strong magnetic coupling of the cold atoms to the cavity mode, and has the advantage of scalability and integration.

This work is supported by the National Nature Science Foundation of China under Grant Nos.60778021 and the National
Key Project of Basic Research Development under Grant No.2010CB923102. P.-B.L. acknowledges the support from the New Staff Research Support Plan of Xi'an Jiaotong University under No.08141015 and
the quite useful discussions with Hong-Yan Li.



\end{document}